# Latent table discovery by semantic relationship extraction between unrelated sets of entity sets of structured data sources

Gowri Shankar Ramaswamy[1] and F Sagayaraj Francis[2]

[1] Department of Computer Science & Engineering, Pondicherry Engineering College,
Puducherry, Puducherry 605014, India

[2] Department of Computer Science & Engineering, Pondicherry Engineering College,
Puducherry, Puducherry 605014, India

**Abstract**

Querying is one of the basic functionality expected from a database system. Query efficiency is adversely affected by increase in the number of participating tables. Also, querying based on syntax largely limits the gamut of queries a database system can process. Syntactic queries rely on the database table structure, which is a cause of concern for large organisations due to incompatibility between heterogeneous systems that store data distributed across geographic locations. Solution to these problems is answered to some extent by moving towards semantic technology by making data and the database meaningful. In doing so, relationship between sets of entity sets will not be limited only to syntactic constraints but would also permit semantic connections nonetheless such relationships may be tacit, intangible and invisible. The goal of this work is to extract such hidden relationships between unrelated sets of entity sets and store them in a tangible form. A few sample cases are provided to vindicate that the proposed work improves querying significantly.

***Keywords:*** *Semantic Database, Semantic Relations, Latent Table Discovery, Connecting unrelated tables, Semantic Querying.*

## 1. Introduction

Database management systems have been in use since time immemorial and their purpose is quite evident. Storing data in an organised manner and retrieving them as and when needed are two major objectives of any database system. Retrieving the stored data is typically using query languages like Structured Query Language (SQL). The formative years of database systems dealt more with maintaining precision and integrity of the data being stored, minimising redundancy so as to conserve space and optimise query processing for faster result delivery. With the advent and growth of the World Wide Web, a new form of retrieval named Information Retrieval came into being. People wanted not just raw data from the storage systems but processed and meaningful information. The burgeoning search engines that provided natural language interfaces for querying instead of the traditional query languages fuelled this demand. To a system designer this is an overwhelming task but to a naive user these interfaces made computing systems user-friendlier. Natural language processing, data mining, text mining and a host of other allied fields collaborated to give answers to the demands of the Internet albeit not extensive enough to satisfy the growing needs. Researchers sought a more powerful way to improve querying and retrieval efficiency that is when Semantic technology was conceptualised. All along, retrieval has been approached only in terms of syntax. Semantic retrieval was a new dimension to information retrieval rather an altogether shift in the way retrieval is viewed. Semantics, as per linguistic definition [1] is - the branch of linguistics that deals with the study of meaning, changes in meaning, and the principles that govern the relationship between sentences or words and their meanings. In the ambit of data management and information systems, semantics also defines the relationship between concepts and amongst instances of the concepts that the data embodies. By establishing meaningful links between concepts, information that is retrieved can be made more meaningful. Also, certain syntactical incompatibilities between heterogeneous systems can be overcome by introducing a semantic layer that is common to both. Much of the development in this field has taken place with unstructured data sources like web pages through natural language processing assisted by ontologies. Structured data sources have been less explored nevertheless they also feed enormous amount of data to the web. Linking data within structured data sources i.e. relational tables relies on referential integrity constraints introduced during the design. But less has been thought about tables that are in no way related by such constraints. This leads us to explore the possibility of identifying semantic relationships from within databases designed hitherto using relational model.

## 2. Related Work

A detailed study of literature revealed various perspectives to linking data and publishing linked data to facilitate faster and efficient querying. However, no methods were found to be exactly in line with our work of extracting hidden relationships. There are various other existing approaches to create linked data and to publish linked data from relational databases. Some of them are discussed briefly.

2.1 Ontology generation from relational database

A method to generate ontology from relational databases is presented in this work. The difficulties in ontology generation from relational database include unclear generation approaches, un-unified ontology languages and so on [2]. So in order to provide unified ontology and improve the quality of ontology generation, this work first extracts database metadata information [3] from relational database using reverse engineering technique [4], then analyses the correspondent relationship between relational database and OWL ontology and presents an ontology generation from relational database [2]. Finally, a prototype tool of the generator, implemented based on Jena in Java development platform and case study demonstrates the feasibility and effectiveness of the approach. MySQL is used as the original database because it provides specific views about the database metadata. This approach can mine the implicit conceptual relationship in relational database, i.e., hierarchy relationship, concept constraints, property constraints and so on.

2.2 Reverse engineering by analyzing HTML forms

This is an approach to reverse engineering relational databases to ontologies. It is based on the idea that semantics of a relational database can be inferred without an explicit analysis of relational schema, tuples and user queries [5]. Rather, these semantics can be extracted by analysing HTML forms, which are the most popular interface to communicate with relational databases for data entry and display on the Web [6]. The semantics are supplemented with the relational schema and user "head knowledge" to build ontology. This approach can be applied to migrating data-intensive Web pages which are usually based on relational databases to the ontology based Semantic Web [6]. The three basic steps are: (i) analysing HTML forms to extract a form model schema that expresses semantics of a relational database behind the forms, (ii) transforming the form model schema into an ontology ("schema transformation"), and (iii) creating ontological instances from data contained in the forms ("data migration"). The main reason for this migration is to make the Web content machine-understandable [7]. Data-intensive Web pages do not exist until they are generated dynamically from relational databases at the time of user requests [8]. Since data is usually represented through HTML forms for displaying to the users (i.e. intended for human consumption only), machines cannot understand and process data-intensive Web pages.

2.3 Semantic link discovery over relational database

This is a framework for online discovery of semantic links from relational data. The framework is based on declarative specification of the linkage requirements the user specifies that allows matching data items in many real-world scenarios [9]. These requirements are translated to queries that can run over the relational data source, potentially using the semantic knowledge to enhance the accuracy of link discovery [10, 11]. This framework lets data publishers discover and publish high-quality links to other data sources easily and could potentially enhance the value of the data in the next generation of web. The framework was implemented on a commercial database engine and the effectiveness of this approach was validated by finding links between a real clinical trials data source and several other data sources. The framework can greatly enhance the process of publishing a high-quality data source with links to other data sources on the web. Furthermore, this framework combined with an existing popular declarative approach for generating RDF data on the web such as those mentioned in [12], can lead to a quick and simple way of publishing an online linked data source with high-quality links.

## 3. Proposed Model

Given a set of unrelated tables and correspondence conditions between attributes, new semantic relations have to be discovered by searching a knowledge-base i.e. Ontology. The discovered relations are again tables per se. Fig. 1 show the conceptual view of the problem at hand.

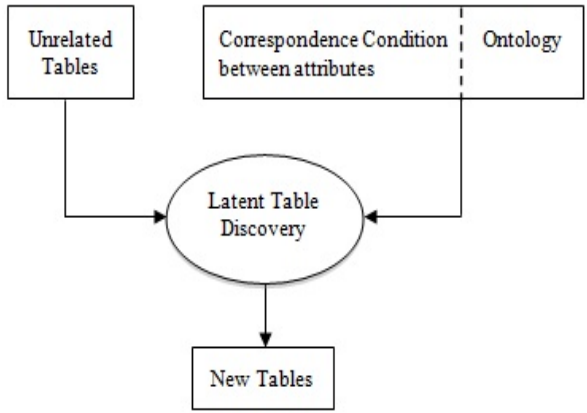

Fig. 1 Conceptual View of LTD

### 3.1 Definition of Latent Table Discovery (LTD)

Given a set of entity sets, LTD discovers other closely related entity sets that are nonexistent in the set of entity sets earlier by taking into account their co-occurrence in domain ontology.

### 3.2 The Latent Table Discovery Method

Given a set of relational tables and a correspondence condition on the columns to be associated, a semantic link can be established by doping an external entity between the attributes. This external entity comes from different possible world to which these attributes would belong to i.e. typically ontology. Based on the given correspondence conditions, the ontology is searched to find a match on both the attributes. If a semantic similarity exists, a semantic link between the two tables via the chosen attributes is said to be present.

In mathematical terms, we have unrelated relational tables with K set of tuples and Y set of attributes. The method to find out the link between these tables is:

1. Input is two such tables $T_1$ and $T_2$ and a correspondence condition between an attribute $Y_1$ of $T_1$ and $Y_2$ of $T_2$.
2. An ontology containing the domain knowledge of all the attributes of the tables represented as $O = \{C, L\}$ where $C$ is a set of concepts and $L$ is a set of links between the concepts is also considered.
3. The knowledge base is traversed to find semantic match with $Y_1$ as well as with $Y_2$ through one or more intermediate concepts. By semantic match we mean a match derived out of thesauruses that point to the concept in contention.
4. A new relation is excavated as shown in Eq. 1

$$Y_1 \xrightarrow{X} Y_2 \quad (1)$$

where X is one or more concepts connecting $Y_1$ with $Y_2$ semantically. Observe that we have made a transitive link between $Y_1$ and $Y_2$ via X. Also, if X is set of concepts then this becomes a transitive closure of the Ontology formally written in Eq. 2

$$Y_1 \xrightarrow{X^*} Y_2 \quad (2)$$

This discovery is thus a semantic transitive closure.

5. The new relations mined by step 4 has three attributes namely $Y_1$, X and $Y_2$. In order to convert this into ontological knowledge i.e. RDF triplets, the following procedure can be followed:
   for each new relation R
     for each tuple T within R
       Map $Y_1$ to rdf:Subject
       Map X  to rdf:Predicate
       Map $Y_2$ to rdf:Object
     end for
   end for

LTD, as it can be observed, uses well-defined mathematical structures (like transitivity) to excavate hidden relationships between entity sets within a relational table (Intra LTD) or between entity sets across relational tables (Inter LTD).

## 4. Case Study

Latent Table Discovery was applied on a medical database containing two tables namely a diagnosis table (Table 1) and a drug table (Table 2). The latent table obtained was found to possess much significance in giving more meaningful results.

Table 1: Diagnosis

| Complaint | Intervention |
|---|---|
| Fever | High Temperature |
| Diabetes | High Blood Sugar |
| Palpitation | High Blood Pressure |
| Giddiness | High Blood Pressure |
| Anaemia | Low Haemoglobin |

Intervention in medical terms is a set of test or procedures conducted before diagnosis.

Table 2: Drug

| Name | Chemical Composition |
|---|---|
| Crocin | p-AminoPhenol |
| Glibenclamide | Sulfonylurea Glibenclamide |
| Amlogard | Amlodipine |
| Dolo Cold | p-AminoPhenol |
| Feosol | Ferrous Sulphate |

The correspondence condition is between Intervention and Composition attributes.

The ontology used is shown in Fig. 2.

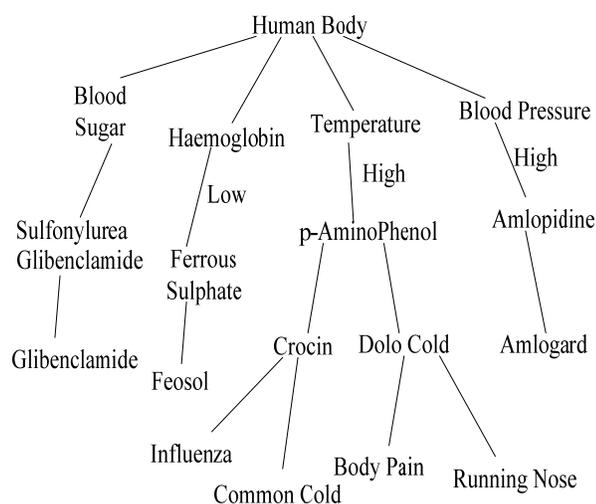

Fig. 2 Ontology of Human Body

The new relation after applying the proposed work will look like:

Table 3: LTD Intervention vs. Drug

| Intervention | Condition | Drug |
|---|---|---|
| Temperature | High | Crocin |
| Temperature | High | Dolo Cold |
| Blood Pressure | High | Amlogard |
| Blood Sugar | High | Glibenclamide |
| Haemoglobin | Low | Feosol |

As it can be seen, High Temperature is associated with Crocin (drug name) whose composition is p-AminoPhenol. Hence if the intervention says temperature is high, it is semantically linked with Crocin.

It can also be inferred that via composition, another similar drug named Dolo Cold is related which in turn points to Influenza meaning that the intervention on temperature gives clue on the other related interventions that have to be performed on the patient. We call such link discovery as semantic transitive closures.

It should also be noted that the new relations again form a table with tuples matching the given correspondence condition thereby extracting a latent table. Also, this table is found to be in the form of RDF triples (Subject, Predicate, Object) where Intervention is the subject, Condition is the predicate and Object is Drug as mentioned in the case study. Hence RDF documents can be auto-generated from this latent table thereby contributing additional knowledge to the existing knowledge base.

## 5. Proof of feasibility of LTD

The crux of this approach is finding semantic relation between concepts.

Let us suppose that we have two concepts X and Y between which a semantic connection is to be established.

Let the binary representation of these concepts take 'n' and 'm' bits respectively.

From set theoretic notations, these two concepts can be visualised as two domains X and Y with cardinalities 'n' and 'm' respectively as shown in Fig. 3.

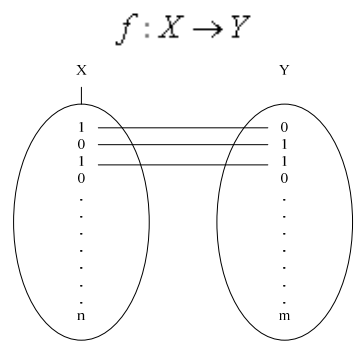

Fig. 3 Concept Mapping in Binary representation

Each of the 'm' images of the co-domain Y can take one of the 'n' possible pre-images of the domain X as a result of the function $f$ which in our case is the semantic relatedness function between the two concepts.

Note that, there can be a number of ways in which this function is written. Estimating the total number of functions that can exist and the complexity of each is typically a problem of computability theory and combinatorics.

According to Church-Turing thesis [13], a computable function [13] is one which can be calculated by a mechanical calculation device like a Turing machine given unlimited amounts of time and storage space. Alternatively, a function is computable if it has an algorithm.

In our case, the function $f$ is computable if there is a Turing machine that takes x as input and halts and returns output $f(x)$ which is a concept Y semantically related to X. Since ontologies are subjective, there may be cases where no semantically related concept exists. If function $f$ produces such results then $f$ is said to be a non-computable function [13].

Case study presented in this paper shows ideal scenarios to exist. Hence $f$ follows the fact that not all functions are computable or non-computable but it is possible produce infinite ideal cases simulating the existence of such sets. These sets are known as recursively enumerable sets [14]. Synonymously, they can be called as recursively enumerable, computably enumerable, semi decidable, provable or Turing recognizable [14]. Hence LTD is theoretically feasible.

## 6. Performance Analysis

Analysis of this problem shows that 'n' elements of the domain can be mapped to 'm' elements of the co-domain in the worst case in $n^m$ ways.

Since the output is dichotomous i.e. presence of a semantic match or absence, the time taken is $O(2^{n^m})$.

If more than two concepts are considered, the complexity is still expected to grow exponentially making LTD a likely case of NP-Complete problem.

## 7. Conclusion & Future Work

The information age is posing tremendous challenges to database researchers with increasing complexity of queries. To cope up with semantic web trend, it is imperative to think about making database systems meaningful and machine understandable in order to facilitate semantic processing. This work envisages an autodidactic database system that can understand the semantics of the data it contains, identify relationships between or amongst data sets, create and update its own knowledge base as well as publish knowledge to the semantic web. Traditionally, databases feed data to the web that is then matched with ontologies to create meaningful content. Nevertheless, this work intends making the database create and store meaningful content and then feed it to the web. The latter has significant improvement over the former. Ontologies, being subjective, can give precarious results. Making the database create its own ontology reduces ambiguity and increases quality of the results. Also, knowledge is periodically mined and updated. Going by this way helps increase interoperability between heterogeneous systems.

Future work is to find a polynomial time reduced solution to LTD.

**Gowri Shankar Ramaswamy** is a post graduate student at Pondicherry Engineering College. His prior occupation was as Systems Engineer developing SOA applications for legacy systems. He has published his previous research paper in Int'l Conference on Web Intelligent Systems (ICWIS 2009) titled "Web based high speed terrain rendering for GIS". His research interests include Geographic Information Systems, Data Semantics.

**Dr. F Sagayaraj Francis** holds a doctorate in Data Management. He is an associate professor at Pondicherry Engineering College. He has to his credit 6 journal and 7 conference publications. His research interests include Data Management, Geographic Information Systems.